# Collective Identity Formation on Instagram – Investigating the Social Movement Fridays for Future

*Research in Progress*

### Felix Brünker
Professional Communication in Electronic Media/Social Media
University of Duisburg-Essen
Duisburg, Germany
Email: felix.bruenker@uni-due.de

### Fabian Deitelhoff
Institute for the Digital Transformation of Application and Living Domains
University of Applied Science and Arts
Dortmund, Germany
Email: fabian.deitelhoff@fh-dortmund.de

### Milad Mirbabaie
Faculty of Business Studies and Economics
University of Bremen
Bremen, Germany
Email: milad.mirbabaie@uni-bremen.de

## Abstract

In recent years, social media changed the way individuals participate in social movements. While activists demonstrate on the street to fight for a public goal, members of specific movements can also act collective online. Thus, different aspects might influence the formation of collective identity and therefore drive collective action on social media. This study combines the perspectives of social identity- and identity theory in order to examine how members of an opinion-based group contribute to the collective group/social identity formation and therefore, to collective action. To this end, we applied automated text classification techniques to Instagram communication related to the social movement Fridays for Future. Analysing 1,137 comments showed that individuals mainly express Group Cohesion and Emotional Attachment rather than Solidarity by commenting on Instagram. This study further presents a proposed model of collective group/social identity of collective action. Succeeding research aims at enhancing the classification and testing the model.

# 1   INTRODUCTION

The way individuals participate in social movements changed over the last decades due to the development of social media platforms (Tye et al. 2018). Individuals now have opportunity structures (Mirbabaie et al. 2016) to be part of social movements in a virtual way. This step breaks the boundaries of physical participation in social movements on the streets. However, individuals can not only be part of a movement by a low-cost effort via social media but rather activist may use social media to catalyse the participation and popularity of certain social movement which they support. In this way, the relevance of social movements may increase for society and political agenda such as the #metoo debate did (Manikonda et al. 2018). However, one recent movement which tries to put the topic of climate change on the local, as well as global political agenda, is *Fridays for Future* (Neimark 2019). To this end, individuals organise and communicate via social media (Stieglitz et al. 2016) and therefore establish online opinion-based groups (Stieglitz et al. 2018; Thomas et al. 2012) by dedicated social media accounts such as the profile of "Greta Thunberg", the public face of Fridays for Future. Further, being part of such an opinion-based group might support common convictions and therefore show up collective actions.

Regarding this context through the lens of identity theory reveals that not only a group membership influences the formation of identity within an opinion-based group. Likewise, information systems such as social media might play a discrete role in identity formation (Carter et al. 2017; Carter and Grover 2015; Liu and Chan 2010). According to identity theory, individuals form their identities by internalised meanings connected to the *Self (Person)*, an individualised *Role*, or as a member of a *Group* (McCall and Simmons 1966; Stets and Serpe 2013). However, identity theory does not comprehensively include inter- and intragroup dynamics (Hogg 2018; Turner et al. 1987). Likewise, regarding identity formation during social movements reveals a multitude of group dynamics which identity theories miss to explain in-depth (Postmes et al. 2006). Therefore, we consider *social identity theory* with its subtheory *collective identity* to fulfil this gap (Davis et al. 2019).

Therefore, we plan to examine how collective action is influenced by the collective group and social identity, i.e. collective group/social identity. As this is research in progress, we aim to answer as a first step how collective group/social identity is formed on social media within a social movement. Thus, we derived the following research question:

**RQ1**: How is the collective group/social identity formed on Instagram within an opinion-based community?

In order to answer the above research question, we analysed comments from two posts on Instagram by Greta Thunberg, the public face of the social movement Fridays for Future. Overall, we manually collected 1,137 comments from February 2019 until July 2019 for further analysis. To better understand collective identity formation and therefore collective action on social media, we provide first insights how members of an opinion-based community contribute to collective identity formation. As most of the research on social media focus on Twitter or Facebook (Bunker et al. 2017; Mirbabaie et al. 2017; Oh et al. 2013; Stieglitz et al. 2017), we decided to examine the case Fridays for Future on Instagram as the social media platform of choice. In this way, we aim to broaden the understanding of social movement research on social media by providing findings according to a platform which differs in the way of usage to Twitter or Facebook. Therefore, we expect to provide new preliminary findings of social movements on social media in the area of information systems research.

The paper is structured as follows. First, we provide an overview of the collective identity formation on social media as well as its role in collective action. Second, we outline our research design by describing the examined case, the data collection, and the data analysis. Furthermore, we present our preliminary findings based on the data analysis in the subsequent section. Last, we provide a conclusion of our outcome and an outlook for a follow-up study and further research.

# 2   BACKGROUND

## 2.1   Collective Identity Formation on Social Media

In order to understand the formation of (social) collective identity and its impact on collective action on social media, it is crucial to investigate the role of the individual and (social) group influences (Thomas et al. 2012). To this end, identity theory describes the individual identity by three overlapping dimensions: *Person*, *Role* and *Group*/Social (Davis et al. 2019; Stets and Serpe 2013). The dimensions Person and Role refer to "identities which are internalized meanings attached to the self as a unique person [as well as] an occupant of a role" (Davis et al. 2019, p. 3). Likewise, group identity describes the membership in an (activist) community whereas social identity is untied from a network and an expression of one's status or lifestyle (Davis et al. 2019).

However, taking only identity theory into account is not sophisticating to describe the complex process of collective identity formation during a social movement. Thus, Davis et al. (2019) integrated the theory of collective identity into the dimension group/social of identity theory and called it *collective group/social identity* (see Figure 1). The scholars described collective identity as a subtheory of social identity theory that pertains to activist identification on social media (Davis et al. 2019). In this context, collective group/social identity closes the gap between identity theory and social identity theory by providing a theoretical foundation which considers the interrelation between individual, interpersonal and group/social in an information system such as social media (Burke and Stryker 2016; Davis et al. 2019; Miller et al. 2016).

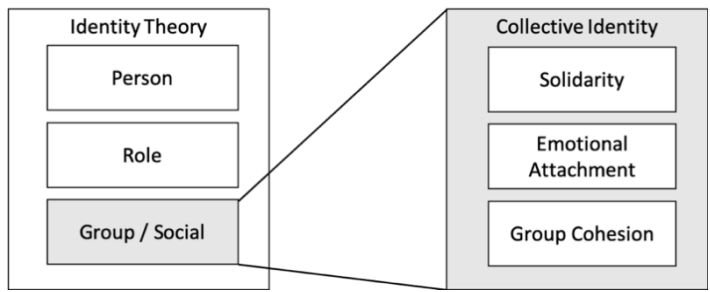

*Figure 1: Conceptual integration of collective identity into identity theory (Davis et al. 2019)*

Furthermore, collective identity combines the bases *group cohesion, emotional attachment,* and *solidarity* (Polletta and Jasper 2001). These three bases might show up on social media in several ways. However, even in heterogeneous networks which are loosely connected are these three bases important for identity formation and maintenance during a social movement (Melucci 1988; Miller et al. 2016). In the context of social movements, social media provides individuals self-verifying capabilities by various participation patterns such as liking, commenting or connecting which others (Miller et al. 2016). Subsequently, evolving personal networks could drive individuals to support the collective. This can be motivated by emotional attachments on social media (Melucci 1988; White 2010). However, several influencing factors might form collective action and drive participation in social movements as the social identity model of collective action (SIMCA) suggests (van Zomeren et al. 2008).

## 2.2 Collective Identity and Collective Action

Collective action can be described as an action which is exclusively based on the voluntary cooperation of individuals (Marwell and Oliver 1993). Furthermore, collective action strives to establish a public or semi-public good (Heckathorn 1996). In this context, existing research often explains group interactions as a central factor for social identity formation (Postmes et al. 2006; Postmes and Spears 2008; van Zomeren et al. 2008) as well as opinion-based group identification (Thomas et al. 2012). According to the integrative meta-analysis of collective action research by van Zomeren et al. (2008), the three subjective variables perceived injustice, perceived efficacy, and social identification could influence collective action. The scholars derived the SIMCA (Figure 2) which emphasises a central role of social identity in promoting collective action. It shows how social identification is formed by group interactions.

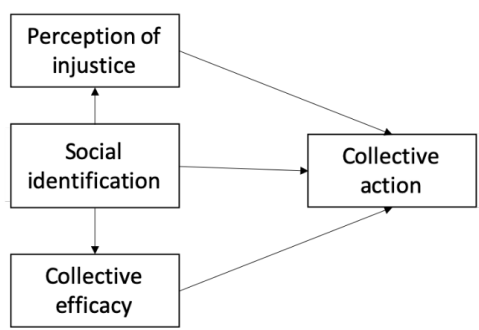

*Figure 2: The social identity model of collective action (SIMCA) (van Zomeren et al. 2008)*

However, Thomas et al. (2012, 2016) proposed an approach focussing on opinion-based group identity as well as group-based effects on the formation of collective action. But, the studies neglected the link between identity theory and the formation of a collective identity towards collective action (Davis et al. 2019). Especially in social movements, individuals forming a collective identity which could emerge on the basis of group members experiencing a common set of interests or goals (Stets and Serpe 2013). However, people differ in the level of participation during a social movement which could be related to individuals' multiple identities or the extent of commitment to the group (Stets and Serpe 2013). Thus, examining the formation of collective identity is crucial to understand the evolving of collective action in social movements. We, therefore, aim to examine collective group/social identity (Davis et al. 2019) to investigate the role of identity formation on collective action.

## 3 RESEARCH DESIGN

### 3.1 Case Description

The social movement Fridays for Future grows in global attention due to the large devotion of the head of the movement, Greta Thunberg. Thunberg is a young Swedish climate activist who started the global phenomenon Fridays for Future by striking school each Friday. Moreover, this social movement is mainly driven by young students. Fridays for Future appeals to the citizen in several countries around the world to stand together and fight against climate change. The members of the movement utilising social media to organise new strikes and demonstrations. Nowadays, the social movement reached global attention. The participation in Fridays for Future scores about 4717 strikes in more than 3800 cities across every continent ("Strike List Countries" 2019).

### 3.2 Data Collection

In order to investigate how collective identity is formed on Instagram by the three bases of collective identity: solidarity, emotional attachment, and group cohesion, we manually collected 1,137 comments of two public available post by Greta Thunberg on Instagram. The two posts were chosen due to their call for global participation on the first (post one) and second (post two) global strike for climate. Furthermore, the social media platform Instagram is chosen as the object of research due to its popularity, especially in the generation of current students. Likewise, public communication about a certain topic builds up under the umbrella of a distinct post on Instagram. Thus, examining the communication related to two major calls for participation in global climate strikes might give relevant insights into the research objective.

### 3.3 Data Analysis

To analyse the collected comments for both posts, we saved the comment data in an SQLite database. For every comment, we collected the publicly available data, consisting of the comment content (text), date of creation, and the username of the poster. Additionally, we connected every comment with the shortcode, Instagram's unique identifier for a post.

For the first analysis step, we annotated every comment with the language used within the comment content. We could not annotate 372 out of the 1,137 comments, because the content contains only mentions of other Instagram users, hashtags, or Emojis. We did not use Emojis, hashtags, or the usernames for deducing the language, because this is not reliable and might lead to false conclusions. We found sixteen different languages used in the comment contents. The dominant language is English (584), followed by Italian (65), Swedish (53), Spanish (25), and French (10). The remaining eleven of the sixteen languages are below ten occurrences each. Therefore, we focused on English comments in our further analysis, to focus on the majority of comments containing one language and its specific characteristics for categorising the comments.

In order to categorise every comment by the three bases of collective identity, we created three lists of words, Emojis, and hashtags. To this end, we randomly selected 200 comments out of the 1,137 and manually categorised words, Emojis, and hashtags to the fitting collective identity base. The subsequent step for categorising the comments into the three collective identity bases is automated through a SQL statement. During this step, a single comment can be assigned to multiple collective identity bases. We searched for the previously defined words, Emojis, and hashtags within the comment contents with a like operator. We do not count the number of occurrences for every word, Emoji or hashtag, we are tracking if one is present or not. Therefore, a comment can be between none or all categories. This is important, because a comment can, for example, be an emotional attachment and enhancing the group cohesion.

## 4 PRELIMINARY FINDINGS

Out of the 584 comments, the above-mentioned SQL statement could not classify 145 English comments in at least one category. This is because of not explainable words (e.g. "periotd"), maybe due to misspelling, very short questions (e.g. "Only ??") or comments in general (e.g. "Me") with just one or two words, or because of text that does not make any sense in the context of Fridays for Future in general or the two posts specifically. Some comments could be classified as troll comments, with content like "lol" or "nah". Such comments can also be an Emotional Attachment, but without more context, we decided to not classify them because the content is ambiguous. For this first analysis, we do not take troll comments or short comments/questions into account. Due to the fact, that these comments are not classified to one of the three bases, they are automatically ignored in our subsequent analysis. The following table shows the words, Emojis, and hashtags for our classification:

| Collective Identity Bases | Words | Hashtags | Emojis |
|---|---|---|---|
| Group Cohesion | we, us, together, strong, can't wait, don't stop, let's do, there, with you, see you, @ | #cleanair, #cleanenergy, #FridaysForFuture, #ClimateStrike, #climatestrikes, #climatecrisis, #SchoolStrike4Climate, #climatechange | - |
| Emotional Attachment | cool, super, yes, stupid, no, fuck, dumb, cunt, asshole, bitch, shit, useless, rock, best | - | 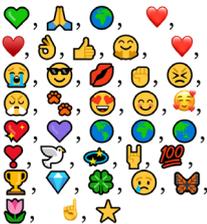 |
| Solidarity | thank, thanks, legend, hero | #GreatGreta, #DropoutGreta, #fightthesystem, #govegan, #endanimalabuse, #vegan, #savetheplanet, #yesshecan | 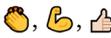 |

*Table 1. Words, Emojis, and hashtags used for classification.*

The corresponding SQL statement categorized the remaining 439 comments as Group Cohesion (299), Emotional Attachment (279), and Solidarity (86). It is evident, that some comments were, at least, categorized as two bases. 24 English comments are in all of the three categories. This is the case for comments with, on average, more text than the other comments (more than 10 words), or due to many used Emojis and/or hashtags. If only comments in exclusively one category are considered, our analysis reveals 120 comments according to Group Cohesion, 100 to Emotional Attachment, and 18 to Solidarity.

The third table shows, that it is uncommon for a comment to belong to just one of our three categories. The Fridays for Future movement arouses strong emotions among their followers. Therefore, and because of the amount of Emojis and hashtags, a comment in the category Group Cohesion is likely to be in the Emotional Attachment category, too, and vice versa. The data and our findings suggest, that the two bases Group Cohesion and Emotional Attachment are the dominant factors expressing collective identity within an opinion-based group in the context of Fridays for Future.

## 5 CONCLUSION AND NEXT STEPS

As this research in progress, we provide preliminary results on collective identity formation on social media. Our findings show to what extent the three bases of collective social/group identity arise on Instagram within an opinion-based community (Friday for Futures). In order to conceptualise the concept of collective identity and collective action, we derived the proposed model of collective group/social identity of collective action (Figure 3). We plan to test this model in a succeeding study to examine the impact of each dimension on the collective group/social identity as well as the indirect effect on collective action. To this end, we aim to conduct a laboratory study based on the preliminary findings of this research in progress. The findings of this study provide first insights to how each base (Solidarity, Emotional Attachment, and Group Cohesion) occur on Instagram within an opinion-based group.

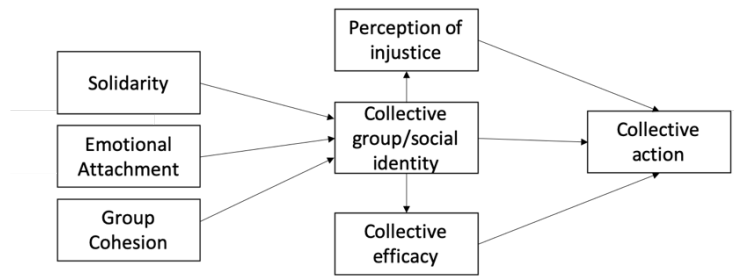

*Figure 3: Proposed model of collective group/social identity of collective action*

However, at this point of research, we need to expand the preliminary findings to observe a broader dataset and review the findings against another sample of data. First, we need further classification on the given dataset, including the phased-out languages. Therefore, we aim to enhance the classification of the comments in order to get more precise results according to the new data. Likewise, we aim to improve the list for the automated classification according to the new data. Second, we plan to gather data from several official Fridays for Future Instagram accounts as well as Twitter communication related to the Fridays for Future movement in order to validate the findings.

**Acknowledgments**

This work was supported by the Deutsche Forschungsgemeinschaft (DFG) under grant No. GRK 2167, Research Training Group "User-Centred Social Media". Furthermore, this project has received funding from the European Union's Horizon 2020 research and innovation programme under the Marie Skłodowska-Curie grant agreement No 823866.